%% file: iclr2020_conference.tex
\documentclass{article} % For LaTeX2e
\usepackage{iclr2020_conference,times}

\input{math_commands.tex}

\usepackage{hyperref}
\usepackage{graphicx}
\usepackage{csvsimple}
\usepackage{url}
\usepackage{float}
\usepackage{subcaption}

\title{Improve robustness of DNN for ECG signal classification: \\ a noise-to-signal ratio perspective}

\author{Linhai Ma \& Liang Liang \\
Department of Computer Science\\
University of Miami\\
Coral Gables, FL 33146, USA \\
\texttt{\{l.ma, liang.liang\}@miami.edu} \\
}

\iclrfinalcopy % Uncomment for camera-ready version, but NOT for submission.
\begin{document}

\maketitle

\begin{abstract}
Electrocardiogram (ECG) is the most widely used diagnostic tool to monitor the condition of the cardiovascular system. Deep neural networks (DNNs), have been developed in many research labs for automatic interpretation of ECG signals to identify potential abnormalities in patient hearts. Studies have shown that given a sufficiently large amount of data, the classification accuracy of DNNs could reach human-expert cardiologist level. A DNN-based automated ECG diagnostic system would be an affordable solution for patients in developing countries where human-expert cardiologist are lacking. However, despite of the excellent performance in classification accuracy, it has been shown that DNNs are highly vulnerable to adversarial attacks: subtle changes in input of a DNN can lead to a wrong classification output with high confidence. Thus, it is challenging and essential to improve adversarial robustness of DNNs for ECG signal classification – a life-critical application. In this work, we proposed to improve DNN robustness from the perspective of noise-to-signal ratio (NSR) and developed two methods to minimize NSR during training process. We evaluated the proposed methods on PhysionNet’s MIT-BIH dataset, and the results show that our proposed methods lead to an enhancement in robustness against PGD adversarial attack and SPSA attack, with a minimal change in accuracy on clean data.
\end{abstract}

\section{Introduction}

Electrocardiogram (ECG) is widely used for monitoring the condition of cardiovascular system. After many years of residency training, a cardiologist becomes experienced in reading ECG graphs, detect abnormalities and classify signals into different disease categories. This is tedious and time-consuming. Researchers found out that deep neural networks (DNNs), especially convolutional neural networks (CNNs) can be trained for ECG signal analysis with excellent classification accuracy \citep{Kachuee2018} \citep{hannun2019cardiologist}. Therefore, with a sufficiently large amount of data and a carefully-designed network structure, DNN models could reach human expert cardiologist level for ECG signal classification, and the analysis for a patient can be done in a fraction of a second. For patients in developing countries where human-expert cardiologists are lacking, a DNN-based automated ECG diagnostic system would be an affordable solution to improve health outcomes.  

However, recent studies have shown that despite the high classification accuracy of DNNs, they are susceptible to adversarial attacks in the form of small perturbations to input of the networks, and the perturbation is even imperceptible to human eyes and not expected (by humans) to change the prediction of DNNs \citep{Akhtar2018}. Adversarial attacks can be classified into two types based on whether the whole structure of the network is known by the attacker. It is a white box attack if the attacker knows the inner structure of the network, such as Fast Gradient Signed Method (FGSM) \citep{Goodfellow2015} and Projected Gradient Descent (PGD)\citep{madry2017towards}. It is a black box attack if the attacker has almost no knowledge of the inner structure of the network, such as transfer-based attack \citep{papernot2017practical} and  SPSA \citep{uesato2018adversarial}. These attacks \citep{Akhtar2018} pose significant threats to the deep learning systems in sensitive and life-critical application fields such as ECG classification. 

To improve DNN robustness, lots of effort has been made by researchers to develop defense methods. Currently, the most popular defense strategy is adversarial training. The basic idea of adversarial training is to added noise to the training samples and the noise is from specific adversarial attacks (e.g. PGD). Through adversarial training, the network can learn some features of adversarial noises and its decision boundary is modified so that it will become difficult to push the input across the decision boundary by adding a small amount of noise. Adversarial training is straightforward but has many problems. For example, generating adversarial samples is very time-consuming, and low-quality adversarial samples can be misleading and even reduce the classification accuracy of networks. Therefore, different adversarial training based defense methods were proposed \citep{Akhtar2018}, which share the same basic idea and vary in how the adversarial samples are generated. Parallel to adversarial training, regularization terms can be added to loss function to reduce the sensitivity of network output with respect to the input. The regularization terms could be the gradient magnitude of loss with respect to input \citep{Ros2017}, or Jacobian regularization \citep{jakubovitz2018improving}. 

\section{Methodology}

In this paper, we proposed two new loss functions with regularization terms to improve robustness of neural networks for ECG signal classification. Our methods aim to reduce the effect of noises added to input, from the perspective of noise-to-signal ratio (NSR), which may make the network more robust. We evaluated our methods on PhysioNet MIT-BIH Arrhythmia ECG dataset \citep{moody2001impact} \citep{ECG}. The results from our experiment show that our proposed methods can achieve a significant improvement in network robustness against PGD white-box attack and SPSA black-box attack, and outperforms the standard adversarial training.

\subsection{ECG Dataset}

The PhysioNet MIT-BIH Arrhythmia ECG Dataset contains 109446 ECG heartbeat records, and there are 5 categories: N (Normal, Left/Right bundle branch block, Atrial escape and Nodal escape), S (Atrial premature, Aberrant atrial premature, Nodal premature and Supra-ventricular premature), V (Premature ventricular contraction and Ventricular escape), F (Fusion of ventricular and normal) and Q (Paced, Fusion of paced and normal and Unclassifiable). The dataset has been divided into a training set (87554 samples) and a testing set (21892 samples), which is publically available \citep{ECG}. We further divided the training set into a 'pure' training set (70043 samples, 80\%) and a validation set (17511 samples, 20\%). The dataset has a large imbalance between classes, and we performed up-sampling to ensure that there are roughly the same number of samples in each class in the training set and testing set.

\subsection{New loss functions to reduce noise-to-signal ratio}

In this section, we will introduce two new loss functions to improve network robustness by reducing noise-to-signal ratio (NSR). The noise refers to the adversarial noise generated from an adversarial attack. The signal refers to the output of the classification network (the logits before the softmax layer). Our methods assume that the nonlinear activation function is ReLU or its variants, and the network may have convolution layers, fully-connected layers, pooling layers, batch-normalization layers, dropout layers, and skip-connections. Since a convolution layer is equivalent to a fully-connected layer with shared weights, we can convert a CNN into a multiple layer perceptron (MLP) in theory. Therefore, we will introduce our method using MLP only in this section. Given an input sample $x$ (a vector), the output of a classification network can be exactly expressed by a "linear" equation \citep{Ding2018}:
\begin{equation}
z=W^Tx+b
\end{equation}
where the weight matrix $W$ will be different for different $x$, and the bias vector $b$ will different for different $x$.  The output is a vector$ z=\left[z_1,\ldots,z_5\right]^T $where $z_i$ is the output logit of Class  $i$ and the total number of classes is 5 in this application. Let $w_i$ be the $i^{th}$ column of $W$ and $b_i$ be the $ i^{th}$ element of $ b$, then we have
\begin{equation}
z_i=w_i^T x +b_i
\end{equation}
The two new loss functions will be introduced in the following sub-sections using the above notations and equations.

\subsubsection{Loss1}

Often, the bias terms of a DNN are very small, and they can be ignored. Therefore, equation (2) is simplified to 
\begin{equation}
z_y = w_y^T x
\end{equation}
where $y$ is the true class label of $x$. $z_y$ is the dot product of two vectors, $w_y$ and $x$. Given the magnitude of $w_y$, the dot product reaches the maximum when the two vectors are aligned in the same direction. In other words, we can train a classification network such that $x$ is "memorized" by $w_y$, similar to a RBF network in some sense (deep RBF network is difficult to train \citep{Goodfellow2015}). Thus, we define a regularization term to encourage the alignment of the two vectors, which is given by 
\begin{equation}
R_1= || w_y - \gamma \frac {x}{x^T x}||_2^2
\end{equation}
where $\gamma$ is a scalar coefficient of the unit vector in the direction of $x$. We combine the regularization term with mean square error (MSE) loss and margin loss for classification, given by
\begin{equation}
loss_1(x,y)=(z_y-1)^2+\sum_{i\neq y} (z_i-0)^2+\sum_{i} max(0,1-z_y+z_i) + \beta_1 R_1
\end{equation}
where $\beta_1$ is a scalar parameter to be determined on validation set. In this way, $\left|z_y\right|=\left|w_y^Tx\right|\rightarrow$1. As a result, $\gamma$ can be fixed to 1. The margin loss and the regularization term $R_1$ are only used for correctly-classified samples; and for wrongly-classified samples, $loss_1$ only contains MSE loss.

This new loss may improve robustness, which can be explained intuitively by Figure 3. Given a small magnitude of adversarial noise $\epsilon$ (a vector), to maximally change the output $z_y$, the direction of $\epsilon$ should be aligned with the direction of $w_y$. Assuming $\epsilon$ and $w_y$ are almost in the same direction, then $\epsilon$ and $x$ will be almost in the same direction because of $loss_1$, i.e., the noise may look like the input. Therefore, if the magnitude of $\epsilon$ is small, it may not alter $z_y$ significantly; and if the magnitude of $\epsilon$ is large enough to cause miss-classification, then $\epsilon$ becomes visually noticeable. 

\subsubsection{Loss2}

Here, we introduce the second loss which directly minimizes noise-to-signal (NSR) ratio. During an adversarial attack, a noisy vector $\epsilon$ is generated and added to the input x, and then the output becomes 
\begin{equation}
z_{y,\epsilon}=w_{y,\epsilon}^T(x+\epsilon)+b_{y,\epsilon}
\end{equation}
If the noise $\epsilon$ is small enough, we can assume that: $w_y\approx w_{y,\epsilon}$ and $b_y\approx b_{y,\epsilon}$. This assumption is valid as long as the "on/off" states of ReLU units do not change significantly and pooling masks do not change significantly when adversarial noise is added. Therefore,
\begin{equation}
z_{y,\epsilon} \approx w_y^T x+b_y+ w_y^T \epsilon = z_y+w_y^T \epsilon
\end{equation}
Then, we define NSR and apply Hölder's inequality:
\begin{equation}
NSR_y=\frac{|w_y^T \epsilon|}{|z_y|} \le \frac{||w_y||_q . ||\epsilon||_p}{|z_y|}
\end{equation}
where $\frac{1}{p}+\frac{1}{q}=1$. In this work, we focus on infinite norm $||\epsilon||_\infty = \epsilon_{max}$, and therefore
\begin{equation}
NSR_y \le \frac{||w_y||_1 . \epsilon_{max}}{|z_y|} = R_2
\end{equation}
We combine the regularization term $R_2$ with mean square error (MSE) loss and margin loss for classification, given by
\begin{equation}
loss_2 (x,y)=(z_y-1)^2+\sum_{i\neq y} (z_i-0)^2+\sum_{i}{max (0,1-z_y+z_i)}+\beta_2 log(1+R_2)
\end{equation}
In the experiment, $\epsilon_{max}$ is set to 1, and $\beta_2$ is determined on validation set.  The margin loss and the regularization term $R_2$ are only used for correctly-classified samples; and for wrongly-classified samples, $loss_2$ only contains MSE loss.

\section{Experiment and Results}

In the experiment, we applied the two proposed losses on two DNNs to evaluate their robustness. One of the networks is the CNN proposed in \citep{Kachuee2018}, which has 30 layers in total. The other network is an MLP designed by us, and it has 8 layers. The structure of this MLP is (187-128)-RELU-(128-128)-RELU-(128-128)-RELU-(128-32)-(32-5). Based on the performance on validation sets, we set $\beta_1$ as 0.2 and set $\beta_2$ as 0.5 to obtain a good balance between robustness and accuracy. Number of epochs is 50, optimizer is "Adamax" and learning rate is 0.001. 

To study the contribution of each term in the loss functions, we also performed ablation experiment. To obtain the baseline performance, the MLP/CNN were trained with Cross-entropy loss for 50 epochs, with "Adamax" optimizer, and learning rate of 0.001. The two DNNs achieved good performance in classification of ECG signals but are vulnerable to adversarial attacks, as shown in figures from Figure 1 to Figure 2. We also evaluated the other two defense methods: (1) Jacobian regularization \citep{jakubovitz2018improving} and (2) standard adversarial training using 10-PGD \citep{shafahi2019adversarial}. 

To evaluate the performance, we compare our proposed methods with other methods, which includes: MLP/CNN with Jacobian regularization \citep{jakubovitz2018improving} and MLP/CNN under 10-PGD standard adversarial training \citep{shafahi2019adversarial} with noise 0.1, 0.2 and 0.3, where 10 is number of steps of PGD attack to generate adversarial samples. In this experiment, there are totally 9 methods to be compared. They are: MLP/CNN with Cross-entropy loss denoted as “ce” in plots, MLP/CNN with mean square error (MSE) loss denoted as “mse” in plots, MLP/CNN with loss1 denoted as “loss1” , MLP/CNN with loss2 denoted as “loss2”, MLP/CNN with Jacobian regularization denoted as “jacob”, MLP/CNN with MSE and margin loss denoted as  “mseMargin” and MLP/CNN trained with adversarial samples generated by 10-PGD adversarial attack under noise level $\epsilon$ denoted as “adv $\epsilon$”. The loss function of this adversarial training is \citep{Goodfellow2015}:
\begin{equation}
loss_{adv} = 0.5 L_{CE}(x, y) + 0.5 L_{CE} (x_\epsilon, y)
\end{equation}
where $L_{CE}$ is Cross-entropy loss and $x_\epsilon$ is adversarial sample. Training batch size is 128 in this experiment. Because we performed up-sampling to ensure that there are roughly the same number of samples in each class in the testing data, average recall is the same as overall accuracy. So, we only report overall accuracy of classification (denoted as "ACC") and average precision across the five classes (denoted as "PREC") as metrics of performance in this experiment. 

%\subsection{20-PGD evaluation}
\subsection{100-PGD evaluation}

First, we used PGD adversarial attack to test these methods. Projected Gradient Descent (PGD)\citep{madry2017towards} is regarded as the strongest first-order attack. K-PGD attacks clean input $x$ with K steps:
\begin{equation}
x^k=\prod\left(x^{k-1}+\alpha\cdot s i g n\left(\nabla_xLoss\left(x^{k-1}\right)\right)\right)
\end{equation}
where $\alpha$ is the step size and $x^k$ is the adversarial example from the $k^{th}$ step. If the noise added to input $x$ is larger than the given noise level $\epsilon$, PGD will project it back to the noise level $\epsilon$. 
\begin{figure}[h]
\begin{center}
\begin{subfigure}{0.45\textwidth}
\includegraphics[width=0.9\linewidth, height=5cm]{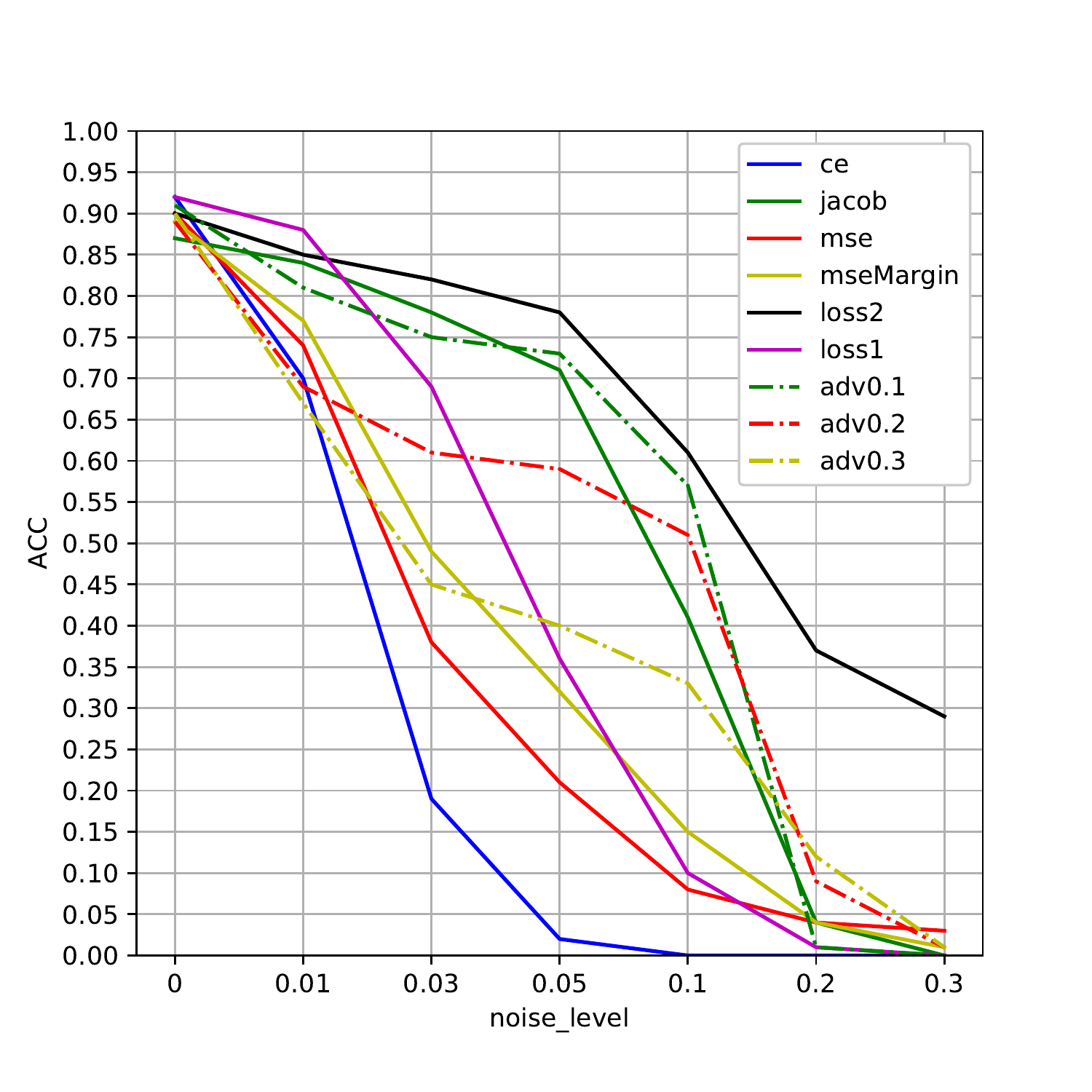} 
\caption{}
\label{fig1-1}
\end{subfigure}
\begin{subfigure}{0.45\textwidth}
\includegraphics[width=0.9\linewidth, height=5cm]{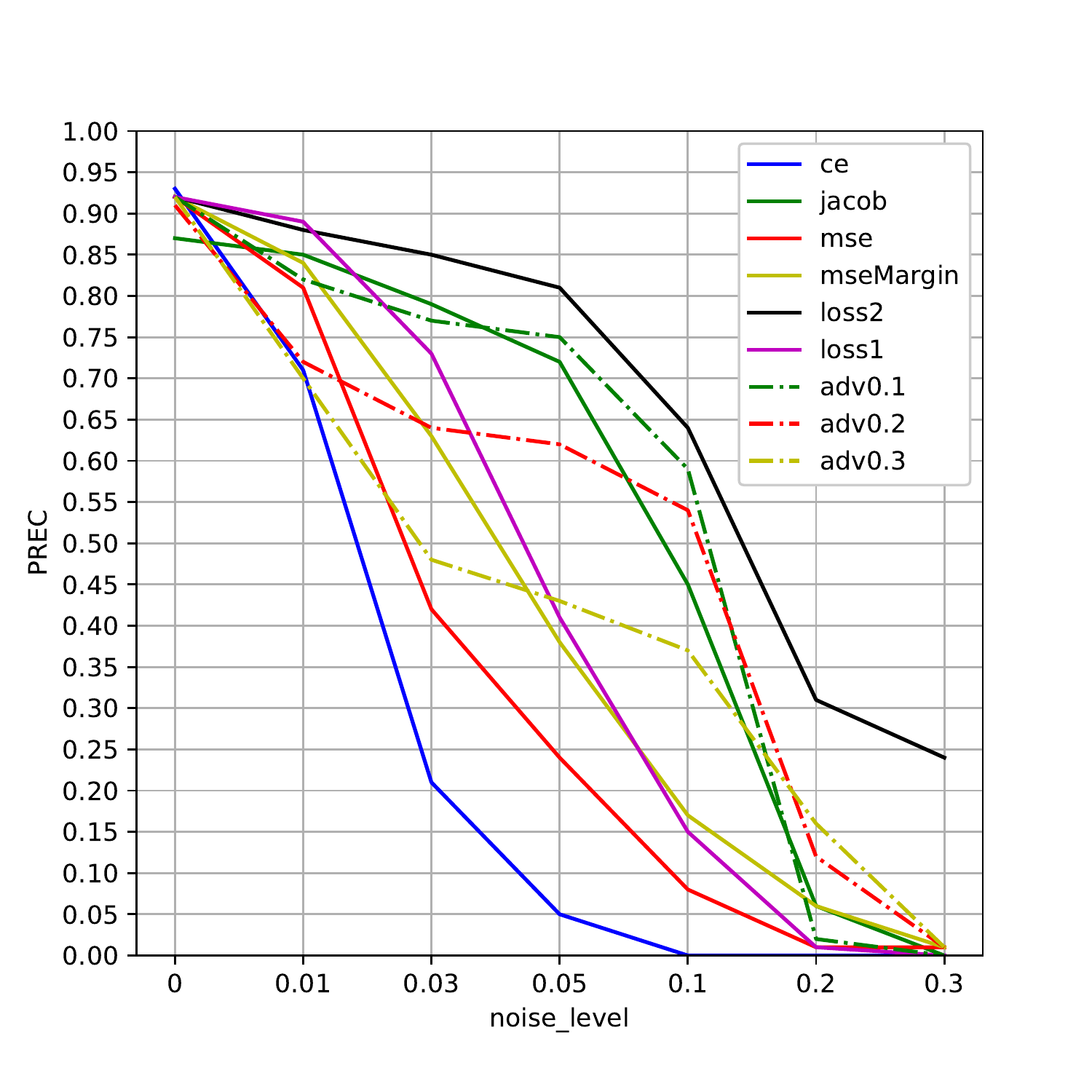}
\caption{}
\label{fig1-2}
\end{subfigure}
\end{center}
\caption{Test accuracy (a) and precision (b) for 100-PGD attack on MLP}
\end{figure}
\begin{figure}[h]
\begin{center}
\begin{subfigure}{0.45\textwidth}
\includegraphics[width=0.9\linewidth, height=5cm]{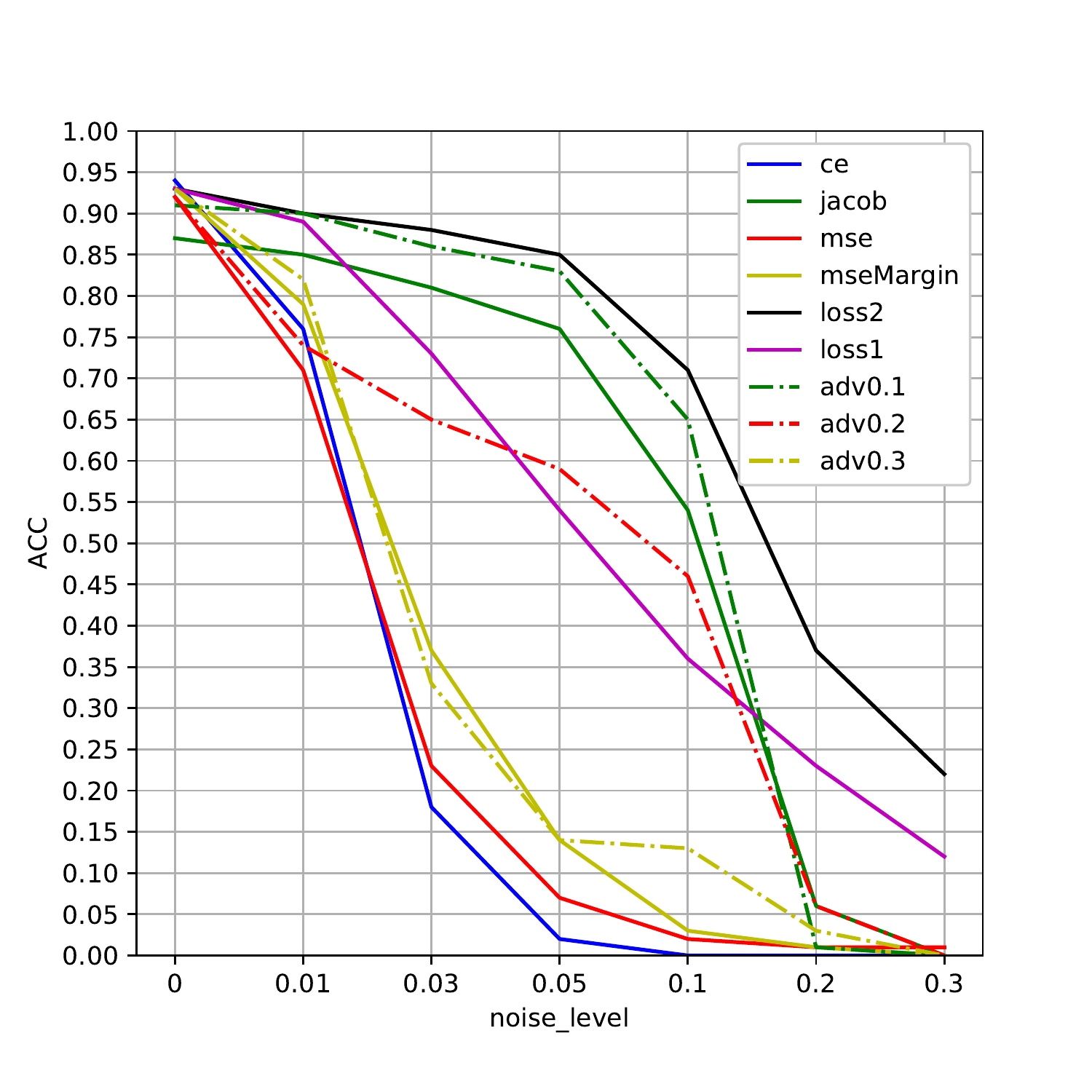} 
\caption{}
\label{fig2-1}
\end{subfigure}
\begin{subfigure}{0.45\textwidth}
\includegraphics[width=0.9\linewidth, height=5cm]{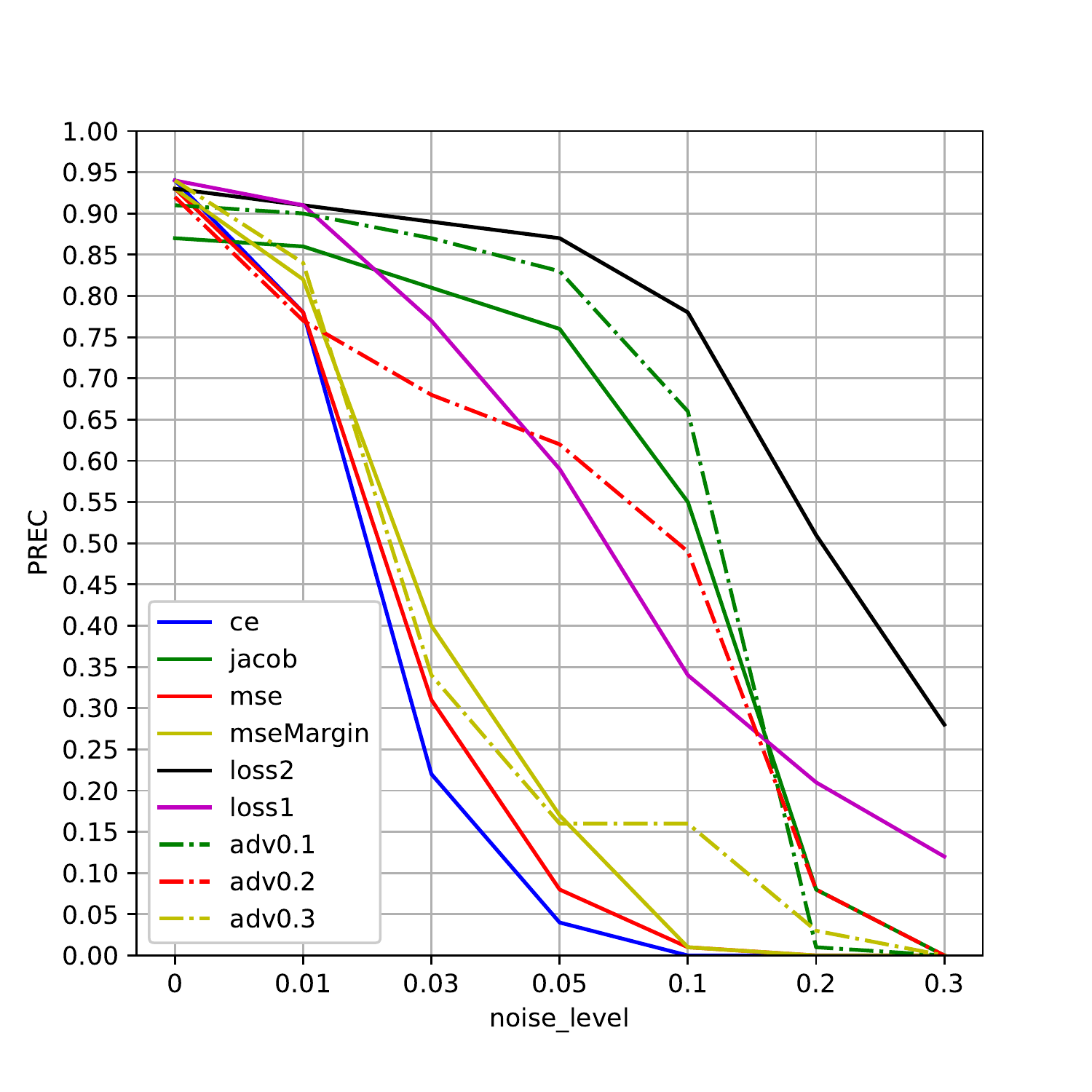}
\caption{}
\label{fig2-2}
\end{subfigure}
\end{center}
\caption{Test accuracy (a) and precision (b) for 100-PGD attack on CNN}
\end{figure}
In this experiment, we used 100-PGD to evaluate the methods. Figure 1 presents the results under 100-PGD attack, comparing our proposed methods to Jacobian regulation and 10-PGD adversarial training for MLP. It can be seen that MLP with our proposed loss2 can achieve the best accuracy and precision under all noise levels except for 0.01 (loss1 is better). Under attack on noise level 0.1, MLP with loss2 has accuracy of 61\% and precision of 64\%, which is the highest among all methods. Figure 2 presents the results under 100-PGD attack, by comparing our proposed methods to Jacobian regulation and 10-PGD adversarial training for CNN. CNN with loss2 has the best performance. Under the PGD attack with the noise level 0.1, CNN with loss2 can maintain an accuracy of more than 71\% and a precision of approximately 78\%. In this experiment, “adv0.2” and “adv0.3” are not as good as loss2 on all of the noise levels, which means these adversarially-trained CNNs are not strong enough to resist against 100-PGD attack. The other tests can be seen in Appendix.

\section{Conclusion}

In this study, we proposed two methods to improve the robustness of deep neural networks for classification of ECG signals. Our methods aim to reduce the proportion of introduced noises in the output of the neural network, namely, noise-to-signal ratio. In this way, our methods can help reduce the effect of noise on the prediction of the network and thus improve the robustness against adversarial attack. The results of the experiment have shown that our proposed loss2, outperforms all other methods under white-box and black-box attacks (PDG and SPSA), for the classification task. We hope that our approaches may faciliate the development of robust and affordable solutions for automated ECG diagnosis for developing countries. 

Note: The implementation of this experiment is available at \url{https://github.com/SarielMa/ICLR2020\_AI4AH}.

\bibliography{iclr2020_conference}
\bibliographystyle{iclr2020_conference}

\appendix
\section{Appendix}

\begin{figure}[h]
\begin{center}
\begin{subfigure}{0.36\textwidth}
\includegraphics[width=0.9\linewidth, height=3cm]{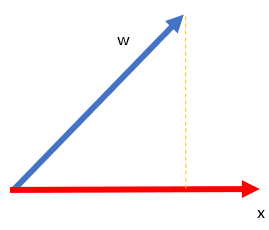} 
\caption{}
\label{fig3-1}
\end{subfigure}
\begin{subfigure}{0.36\textwidth}
\includegraphics[width=0.9\linewidth, height=3cm]{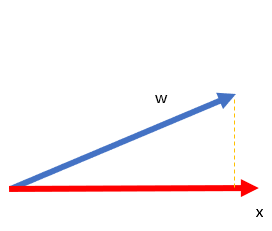}
\caption{}
\label{fig3-2}
\end{subfigure}
\end{center}
\caption{Visualization of 2-dimensional vector $x$ and $w$. (a) shows the condition where $w$ and $x$ are in very different direction while (b) shows the situation where $w$ and $x$ are in more similar direction.}
\end{figure}

\section{Appendix}

In this experiment, we used 20-PGD to evaluate all methods. Figure 4 presents the results comparing our proposed methods to Jacobian regulation and 10-PGD adversarial training for MLP under 20-PGD attack. First, MLP with our proposed loss2 achieved the best accuracy and precision in general. Under attack on noise level 0.1, MLP with loss2 can still have an accuracy 69\% and precision 75\%, which is the highest. Second, performance of this adversarial training is not as good as loss2. We can see that there is an intersection of “adv0.1” and “adv0.2” around noise level 0.1 and an intersection of “adv0.2” and “adv0.3” around noise level 0.2. This means that model trained from adversarial samples under noise $\epsilon$ only behaves well around noise level $\epsilon$. Then, we can notice that both our proposed losses can improve the robustness of MLP against PGD attack while keeping a high accuracy and precision in classification of clean data. MLP with loss1 have accuracy 92\% and precision 92\% and MLP with loss2 can achieve accuracy 90\% and precision 92\%. MLP with “ce” has accuracy 92\% and precision 93\%, and the performance dropped quickly as noise level increased

Figure 5 presents the results comparing our proposed methods to Jacobian regulation and 10-PGD adversarial training for CNN under 20-PGD attack. CNN with loss2 has the best performance as well. Under the PGD attack on noise level 0.1, CNN with loss2 can still maintain an accuracy of 75\% and a precision of 81\%.  We can see that “adv0.2” outperforms “loss2” around only noise level 0.2 and “adv0.3”, and outperforms “loss2” around only noise level 0.3, while on lower noise level it has a very weak resist to PGD attack (shown in Figure 3 (b)). This also demonstrates that model trained with adversarial samples with noise level $\epsilon$ only behaves well around noise level $\epsilon$, which is a weakness of the standard adversarial training. 

\begin{figure}[h]
\begin{center}
\begin{subfigure}{0.45\textwidth}
\includegraphics[width=0.9\linewidth, height=5cm]{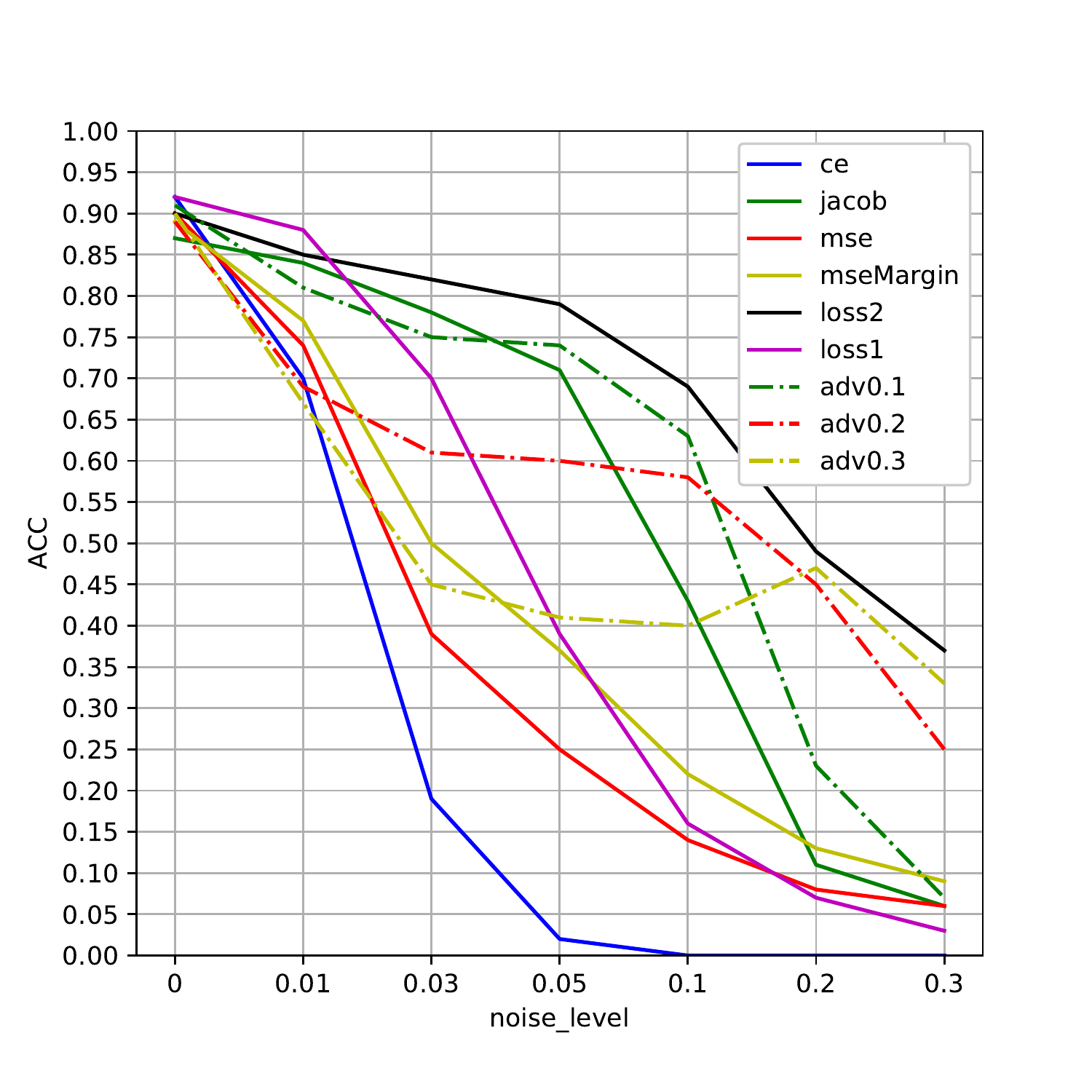} 
\caption{}
\label{fig4-1}
\end{subfigure}
\begin{subfigure}{0.45\textwidth}
\includegraphics[width=0.9\linewidth, height=5cm]{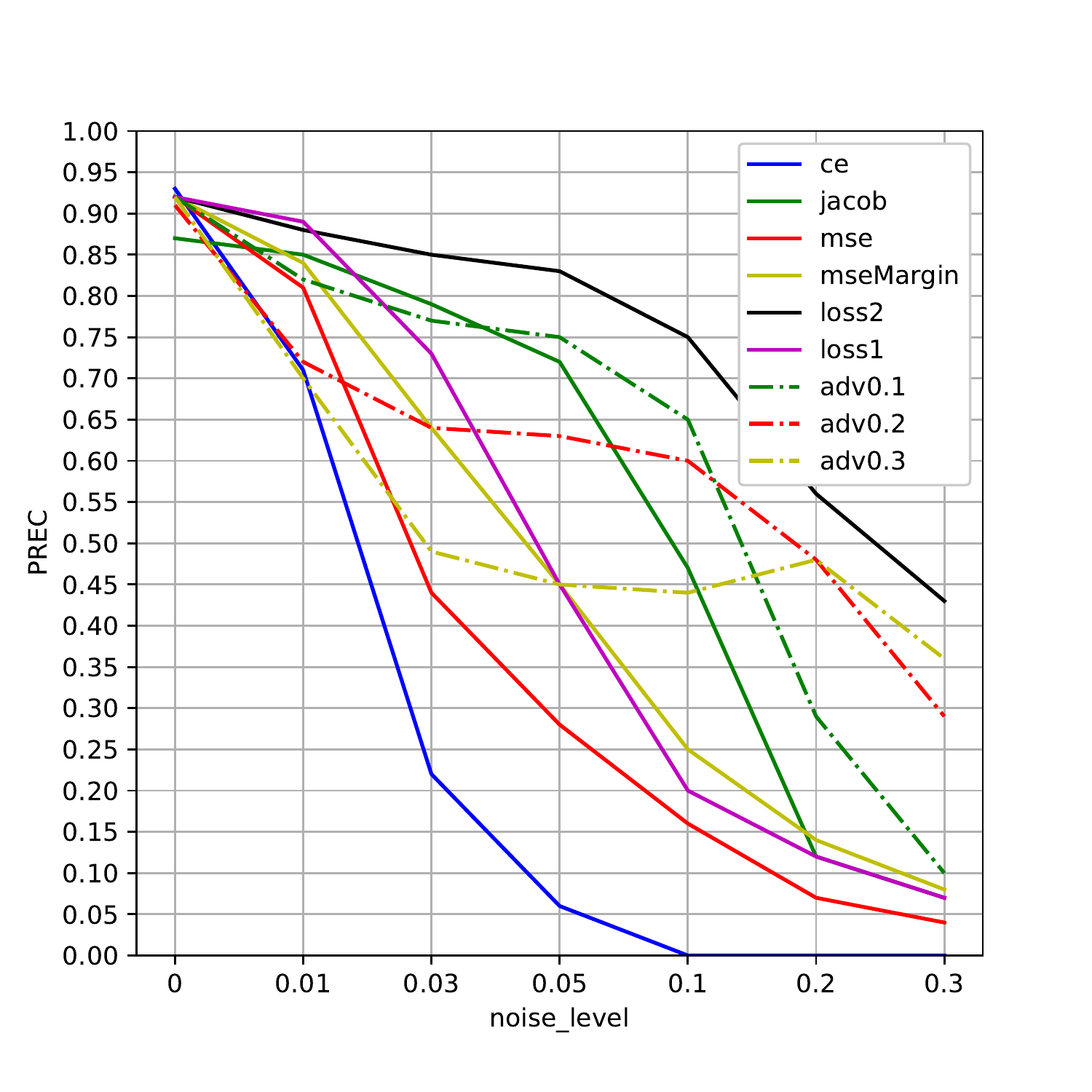}
\caption{}
\label{fig4-2}
\end{subfigure}
\end{center}
\caption{Test accuracy (a) and precision (b) for 20-PGD attack on MLP}
\end{figure}

\begin{figure}[h]
\begin{center}
\begin{subfigure}{0.45\textwidth}
\includegraphics[width=0.9\linewidth, height=5cm]{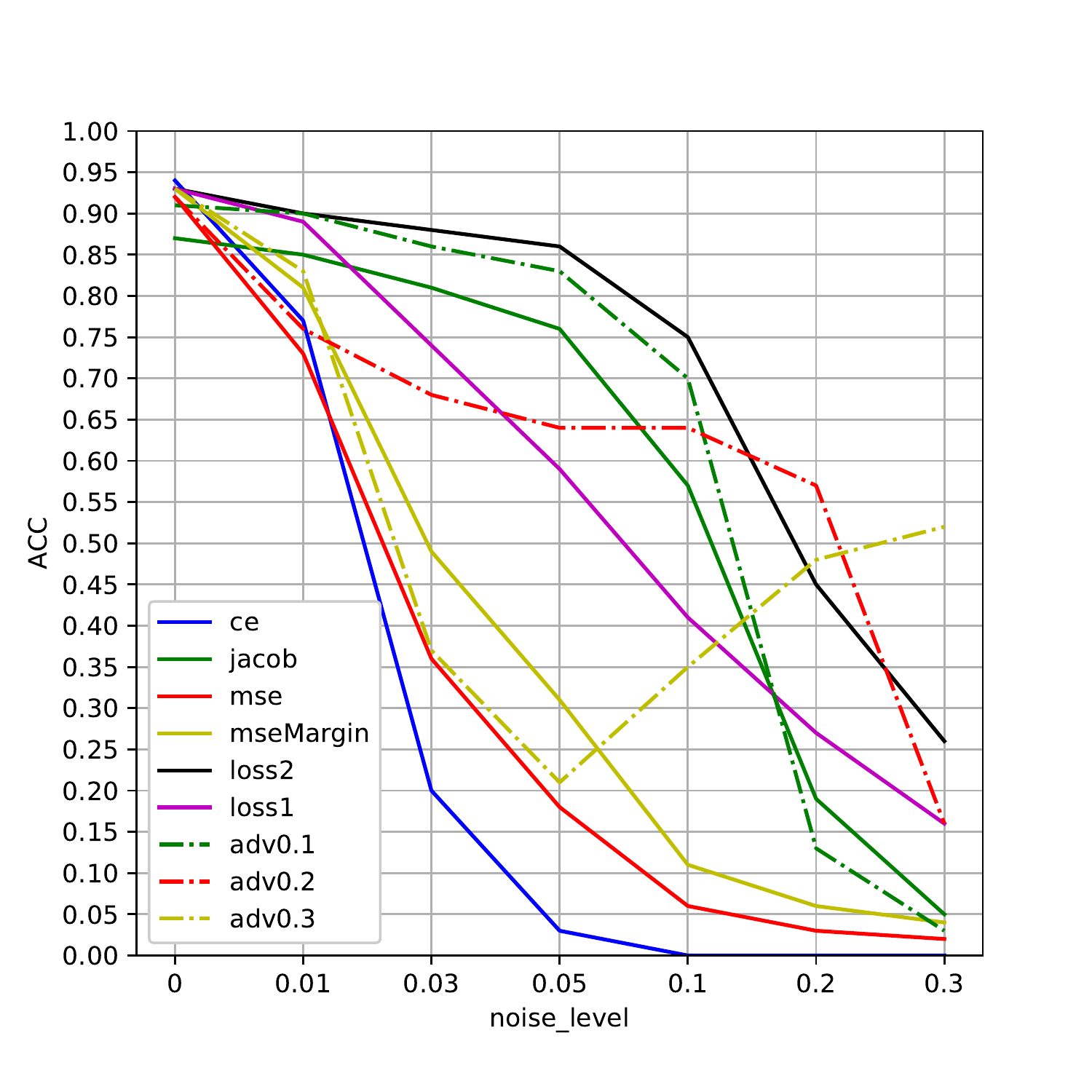} 
\caption{}
\label{fig5-1}
\end{subfigure}
\begin{subfigure}{0.45\textwidth}
\includegraphics[width=0.9\linewidth, height=5cm]{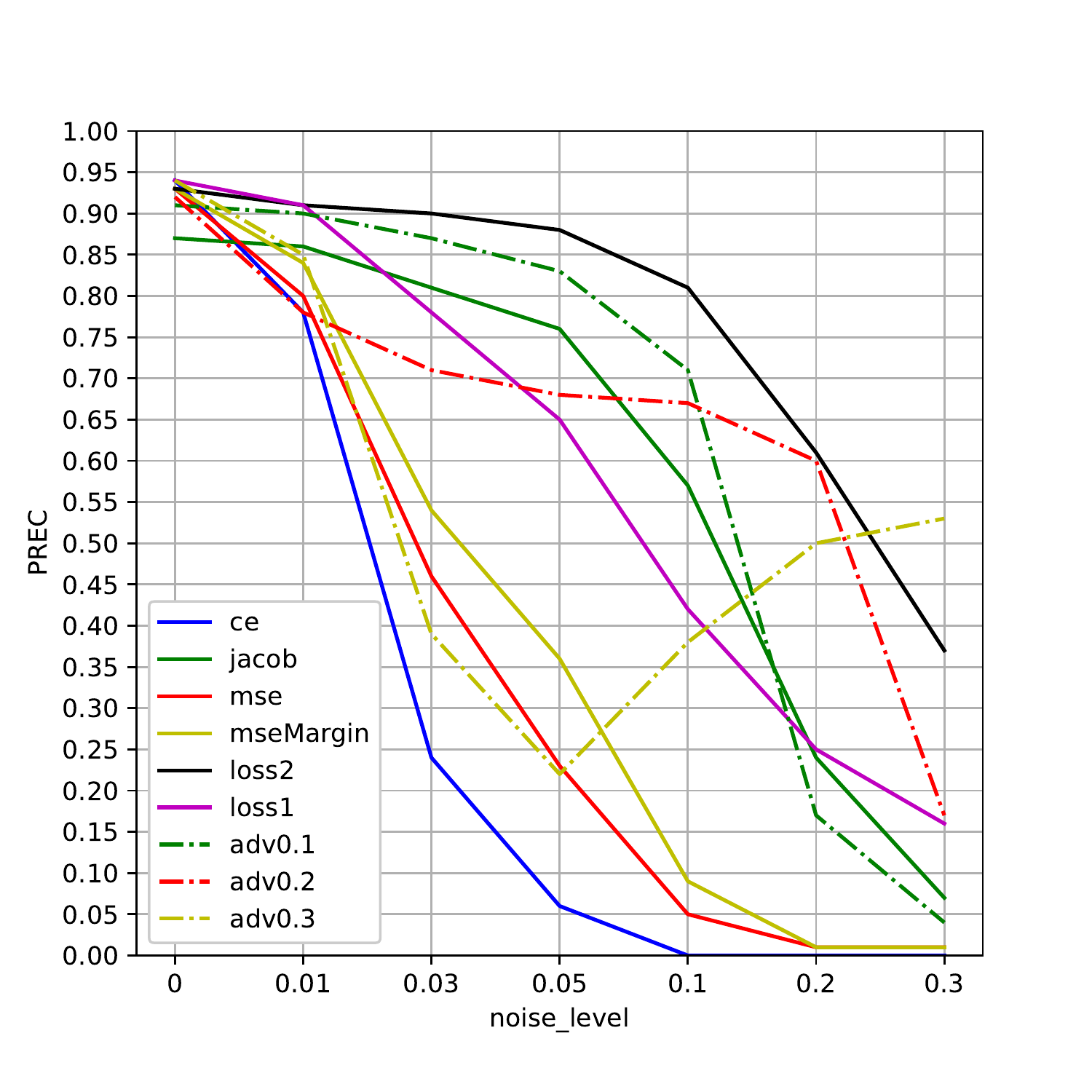}
\caption{}
\label{fig5-2}
\end{subfigure}
\end{center}
\caption{Test accuracy (a) and precision (b) for 20-PGD attack on CNN}
\end{figure}

\section{Appendix}

We evaluated the proposed methods with a black-box attack, SPSA adversarial attack\citep{uesato2018adversarial}. SPSA is a gradient-free method, which is useful when the model is non-differentiable, or more generally, the gradients do not point in useful directions. This attack is applied with almost no knowledge of the inner structure of the network. It randomly generates samples around a given input sample (e.g. an image) of the target network, uses these random samples to estimate the gradient of the target network, and then the estimated gradient can be used for attacks. SPSA runs for 100 iterations with perturbation size 0.01, Adamax learning rate 0.01, and 2048 samples for each gradient estimation.  SPSA attack is very computationally expensive, and therefore we only selected the first 162 samples from each class for this experiment. 

From Figure 6, we can see that loss2 has the best performance in general for the test of MLP. In Figure 6 (a), “jacob” is the best from noise level 0.03 to noise level 0.05, slightly better than that of loss2. However, “jacob” falls sharply when the noise level is larger than 0.05. When noise level is 0.1, loss2 can still have an accuracy of about 51\% while “jacob” method can only achieve 42\%. As the noise level becomes higher, "jacob" drops below 10\%, while loss2 is always above 20\%. In Figure6 (b), we can see that loss2 outperforms “jacob” method under all noise levels. So, loss2 has a better resist against SPSA black-box attack than Jacobian regularization for MLP. In Figure 6 (a), “adv0.1” only outperforms loss2 on noise levels from 0.05 to 0.1, which again confirms that model trained with adversarial samples with noise level $\epsilon$ only behaves well around noise level $\epsilon$. In general, our proposed loss2 has a better performance against SPSA attack than all the other methods in comparing. 

From Figure 7, we can see that under SPSA black-box attack, loss2 has the best performance in general. In Figure 7 (a), “jacob” method is outperformed by loss2. When noise level is 0.1, loss2 can still have an accuracy of about 60\%, better than “jacob” method that can achieve only 52\%. In Figure 7 (b), we can see that loss2 outperforms “jacob” methods almost on all noise levels, with precision of 74\% at noise level 0.1. So, in the test on CNN, loss2 has a better resist against SPSA black-box attack than Jacobian regularization as well. In Figure 7 (a), “adv0.1” outperforms loss2 at noise level only from 0.01 to 0.1, and “adv0.2” outperforms loss2 only on noise levels from 0.1 to 0.2, which again confirms that model trained with adversarial samples with noise level $\epsilon$ only behaves well around noise level $\epsilon$. In general, our proposed loss2 has a better performance against SPSA attack than the other methods.

The good performance of our proposed loss2, under the black-box adversarial attack, also suggests that loss2 is not doing gradient obfuscation to improve the robustness of the target neural network.

\begin{figure}[h]
\begin{center}
\begin{subfigure}{0.45\textwidth}
\includegraphics[width=0.9\linewidth, height=5cm]{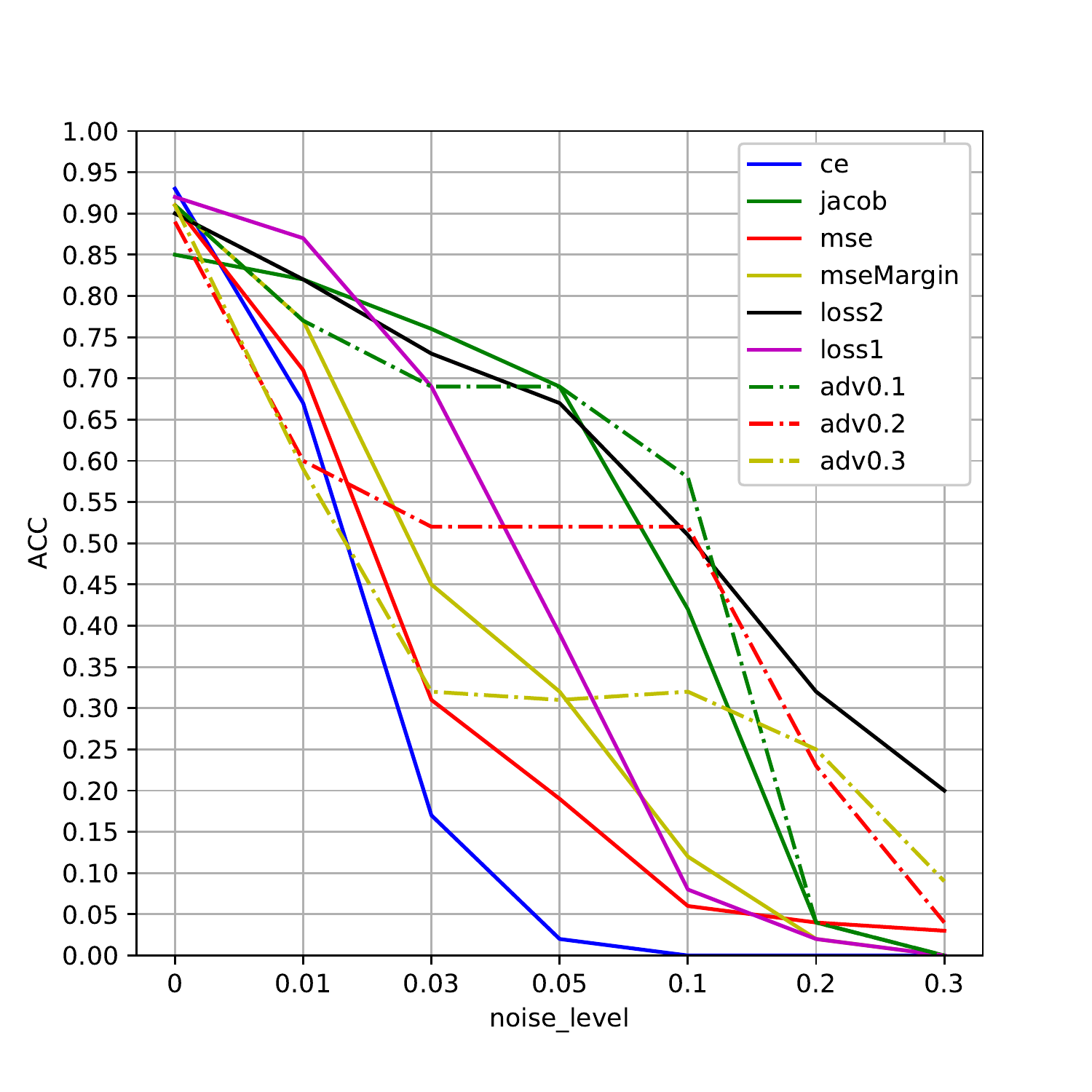} 
\caption{}
\label{fig6-1}
\end{subfigure}
\begin{subfigure}{0.45\textwidth}
\includegraphics[width=0.9\linewidth, height=5cm]{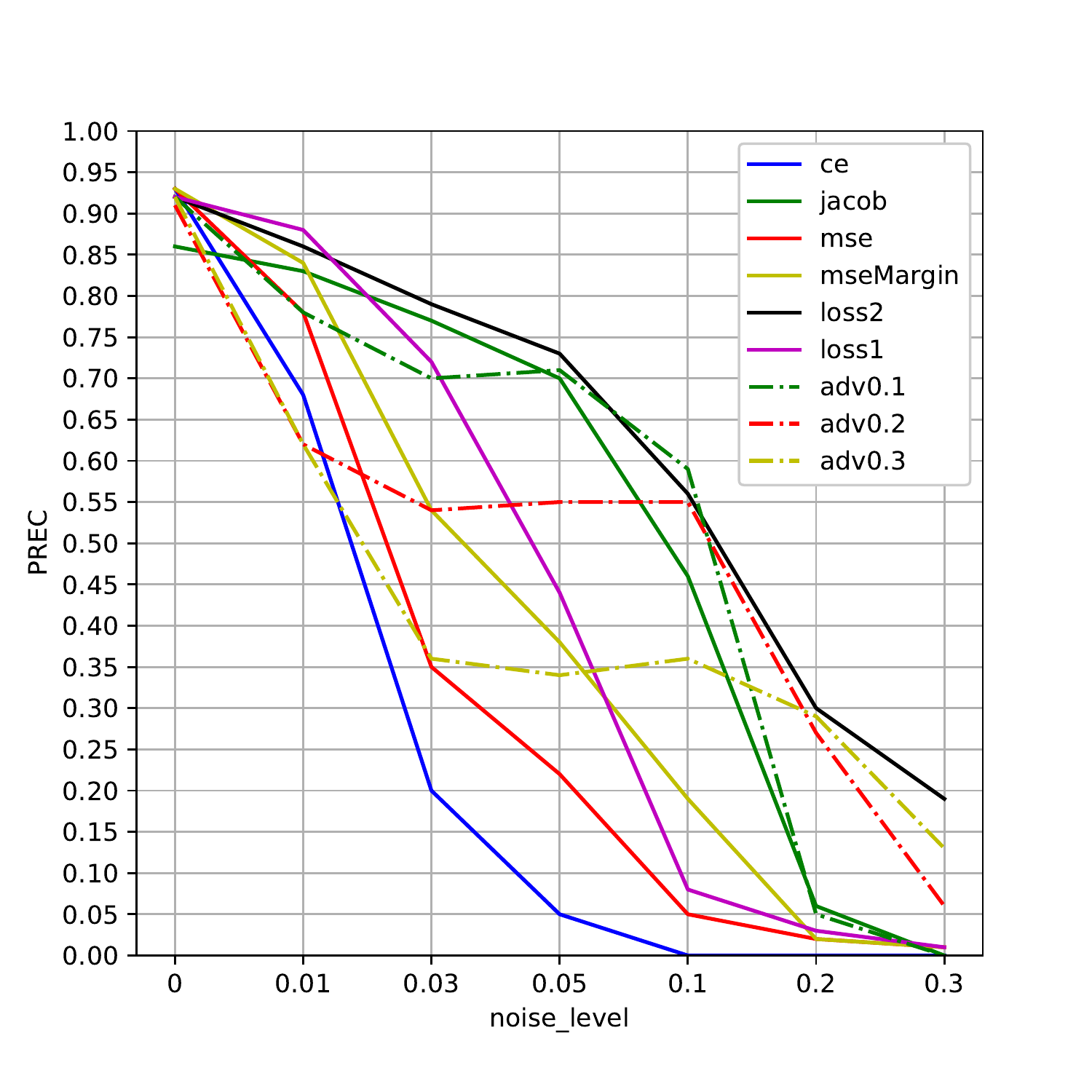}
\caption{}
\label{fig6-2}
\end{subfigure}
\end{center}
\caption{Test accuracy (a) and precision (b) for SPSA attack on MLP}
\end{figure}

\begin{figure}[h]
\begin{center}
\begin{subfigure}{0.45\textwidth}
\includegraphics[width=0.9\linewidth, height=5cm]{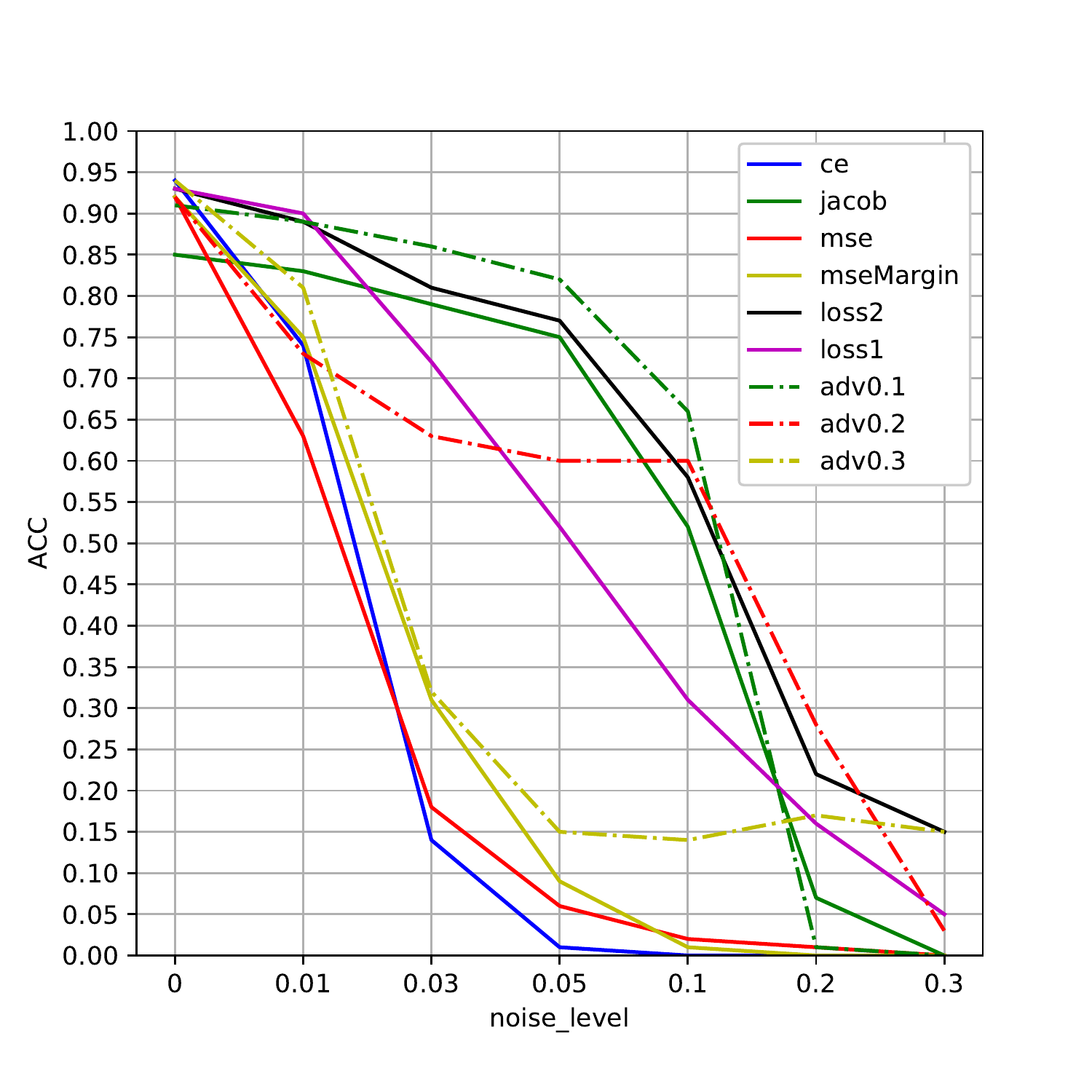} 
\caption{}
\label{fig7-1}
\end{subfigure}
\begin{subfigure}{0.45\textwidth}
\includegraphics[width=0.9\linewidth, height=5cm]{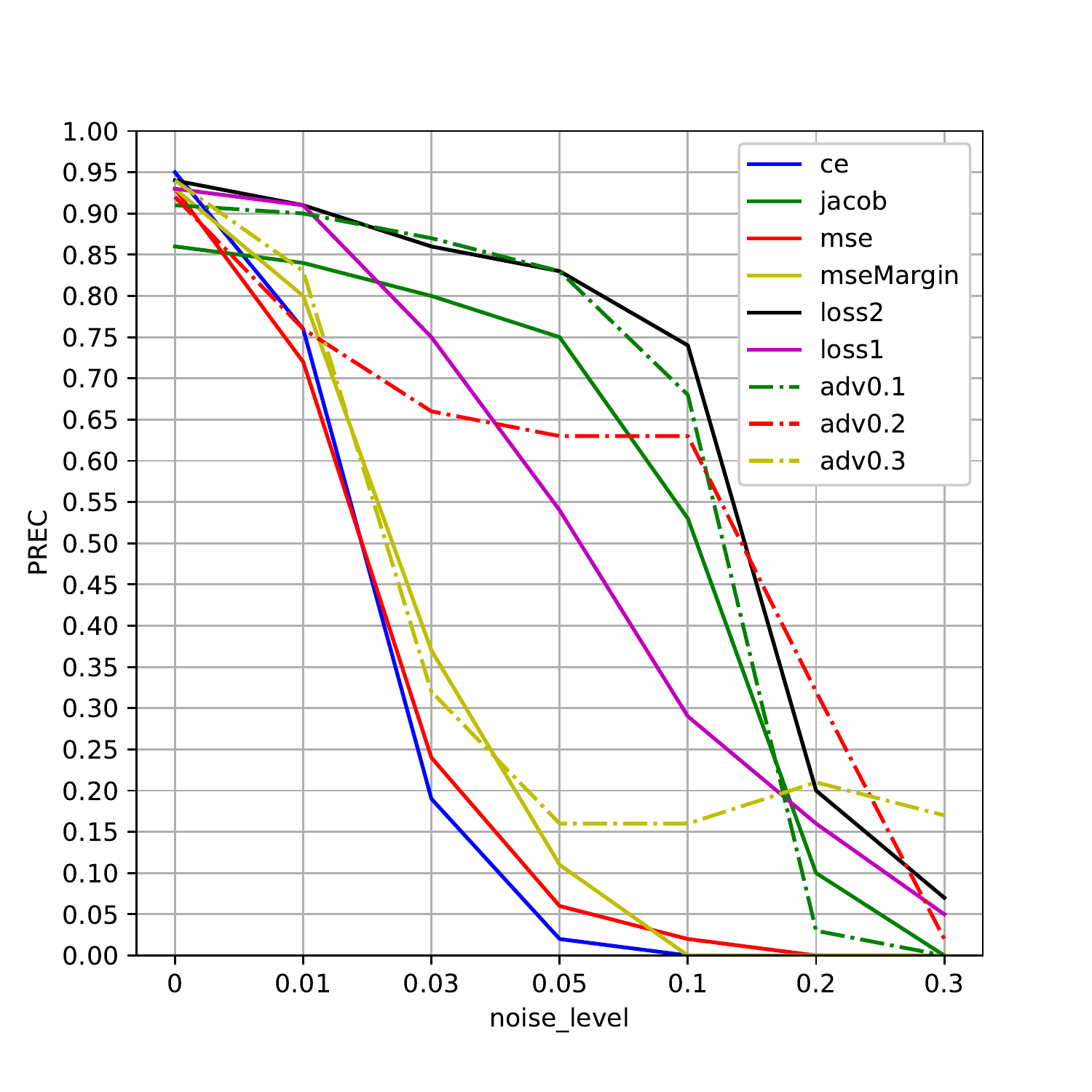}
\caption{}
\label{fig7-2}
\end{subfigure}
\end{center}
\caption{Test accuracy (a) and precision (b) for SPSA attack on CNN}
\end{figure}

\section{Appendix}

The tables corresponding to figures from Figure 2 to Figure 7 are shown in this section.

%MLP 20pgd
\begin{table}[H]
\caption{ACC MLP 20-PGD}
\label{ACC MLP120pgd}
\begin{center}
\begin{filecontents*}{ACCMLP120pgd.csv}
noise level,0,0.01,0.03,0.05,0.1,0.2,0.3

ce,0.92,0.7,0.19,0.02,0.0,0.0,0.0

jacob,0.87,0.84,0.78,0.71,0.43,0.11,0.06

mse,0.9,0.74,0.39,0.25,0.14,0.08,0.06

mseMargin,0.89,0.77,0.5,0.37,0.22,0.13,0.09

loss2,0.9,0.85,0.82,0.79,0.69,0.49,0.37

loss1,0.92,0.88,0.7,0.39,0.16,0.07,0.03

adv0.1,0.91,0.81,0.75,0.74,0.63,0.23,0.07

adv0.2,0.89,0.69,0.61,0.6,0.58,0.45,0.25

adv0.3,0.9,0.67,0.45,0.41,0.4,0.47,0.33

\end{filecontents*}
\csvautotabular{ACCMLP120pgd.csv}
\end{center}
\end{table}

\begin{table}[H]
\caption{PREC MLP 20-PGD}
\label{PREC MLP1  20pgd}
\begin{center}
\begin{filecontents*}{PRECMLP120pgd.csv}
noise level,0,0.01,0.03,0.05,0.1,0.2,0.3

ce,0.93,0.71,0.22,0.06,0.0,0.0,0.0

jacob,0.87,0.85,0.79,0.72,0.47,0.12,0.07

mse,0.92,0.81,0.44,0.28,0.16,0.07,0.04

mseMargin,0.92,0.84,0.64,0.45,0.25,0.14,0.08

loss2,0.92,0.88,0.85,0.83,0.75,0.56,0.43

loss1,0.92,0.89,0.73,0.45,0.2,0.12,0.07

adv0.1,0.92,0.82,0.77,0.75,0.65,0.29,0.1

adv0.2,0.91,0.72,0.64,0.63,0.6,0.48,0.29

adv0.3,0.92,0.7,0.49,0.45,0.44,0.48,0.36

\end{filecontents*}
\csvautotabular{PRECMLP120pgd.csv}
\end{center}
\end{table}

% CNN 20pgd
\begin{table}[H]
\caption{ACC CNN 20-PGD}
\label{ACC CNN1 20pgd}
\begin{center}
\begin{filecontents*}{ACCCNN120pgd.csv}
noise level,0,0.01,0.03,0.05,0.1,0.2,0.3

ce,0.94,0.77,0.2,0.03,0.0,0.0,0.0

jacob,0.87,0.85,0.81,0.76,0.57,0.19,0.05

mse,0.92,0.73,0.36,0.18,0.06,0.03,0.02

mseMargin,0.93,0.81,0.49,0.31,0.11,0.06,0.04

loss2,0.93,0.9,0.88,0.86,0.75,0.45,0.26

loss1,0.93,0.89,0.74,0.59,0.41,0.27,0.16

adv0.1,0.91,0.9,0.86,0.83,0.7,0.13,0.03

adv0.2,0.92,0.76,0.68,0.64,0.64,0.57,0.16

adv0.3,0.93,0.83,0.37,0.21,0.35,0.48,0.52

\end{filecontents*}
\csvautotabular{ACCCNN120pgd.csv}
\end{center}
\end{table}

\begin{table}[H]
\caption{PREC CNN 20-PGD}
\label{PREC CNN1  20pgd}
\begin{center}
\begin{filecontents*}{PRECCNN120pgd.csv}
noise level,0,0.01,0.03,0.05,0.1,0.2,0.3

ce,0.94,0.78,0.24,0.06,0.0,0.0,0.0

jacob,0.87,0.86,0.81,0.76,0.57,0.24,0.07

mse,0.93,0.8,0.46,0.23,0.05,0.01,0.01

mseMargin,0.93,0.84,0.54,0.36,0.09,0.01,0.01

loss2,0.93,0.91,0.9,0.88,0.81,0.61,0.37

loss1,0.94,0.91,0.78,0.65,0.42,0.25,0.16

adv0.1,0.91,0.9,0.87,0.83,0.71,0.17,0.04

adv0.2,0.92,0.78,0.71,0.68,0.67,0.6,0.17

adv0.3,0.94,0.85,0.39,0.22,0.38,0.5,0.53

\end{filecontents*}
\csvautotabular{PRECCNN120pgd.csv}
\end{center}
\end{table}

%MLP 100pgd
\begin{table}[H]
\caption{ACC MLP 100-PGD}
\label{ACC MLP1100pgd}
\begin{center}
\begin{filecontents*}{ACCMLP1100pgd.csv}
noise level,0,0.01,0.03,0.05,0.1,0.2,0.3

ce,0.92,0.7,0.19,0.02,0.0,0.0,0.0

jacob,0.87,0.84,0.78,0.71,0.41,0.04,0.0

mse,0.9,0.74,0.38,0.21,0.08,0.04,0.03

mseMargin,0.89,0.77,0.49,0.32,0.15,0.04,0.01

loss2,0.9,0.85,0.82,0.78,0.61,0.37,0.29

loss1,0.92,0.88,0.69,0.36,0.1,0.01,0.0

adv0.1,0.91,0.81,0.75,0.73,0.57,0.01,0.0

adv0.2,0.89,0.69,0.61,0.59,0.51,0.09,0.01

adv0.3,0.9,0.67,0.45,0.4,0.33,0.12,0.01

\end{filecontents*}
\csvautotabular{ACCMLP1100pgd.csv}
\end{center}
\end{table}

\begin{table}[H]
\caption{PREC MLP 100-PGD}
\label{PREC MLP1  100pgd}
\begin{center}
\begin{filecontents*}{PRECMLP1100pgd.csv}
noise level,0,0.01,0.03,0.05,0.1,0.2,0.3

ce,0.93,0.71,0.21,0.05,0.0,0.0,0.0

jacob,0.87,0.85,0.79,0.72,0.45,0.06,0.0

mse,0.92,0.81,0.42,0.24,0.08,0.01,0.01

mseMargin,0.92,0.84,0.63,0.38,0.17,0.06,0.01

loss2,0.92,0.88,0.85,0.81,0.64,0.31,0.24

loss1,0.92,0.89,0.73,0.41,0.15,0.01,0.0

adv0.1,0.92,0.82,0.77,0.75,0.59,0.02,0.0

adv0.2,0.91,0.72,0.64,0.62,0.54,0.12,0.01

adv0.3,0.92,0.7,0.48,0.43,0.37,0.16,0.01

\end{filecontents*}
\csvautotabular{PRECMLP1100pgd.csv}
\end{center}
\end{table}

% CNN1 100pgd
\begin{table}[H]
\caption{ACC CNN 100-PGD}
\label{ACC CNN1 100pgd}
\begin{center}
\begin{filecontents*}{ACCCNN1100pgd.csv}
noise level,0,0.01,0.03,0.05,0.1,0.2,0.3

ce,0.94,0.76,0.18,0.02,0.0,0.0,0.0

jacob,0.87,0.85,0.81,0.76,0.54,0.06,0.0

mse,0.92,0.71,0.23,0.07,0.02,0.01,0.01

mseMargin,0.93,0.79,0.37,0.14,0.03,0.01,0.0

loss2,0.93,0.9,0.88,0.85,0.71,0.37,0.22

loss1,0.93,0.89,0.73,0.54,0.36,0.23,0.12

adv0.1,0.91,0.9,0.86,0.83,0.65,0.01,0.0

adv0.2,0.92,0.74,0.65,0.59,0.46,0.06,0.0

adv0.3,0.93,0.82,0.33,0.14,0.13,0.03,0.0

\end{filecontents*}
\csvautotabular{ACCCNN1100pgd.csv}
\end{center}
\end{table}

\begin{table}[H]
\caption{PREC CNN 100-PGD}
\label{PREC CNN1  100pgd}
\begin{center}
\begin{filecontents*}{PRECCNN1100pgd.csv}
noise level,0,0.01,0.03,0.05,0.1,0.2,0.3

ce,0.94,0.78,0.22,0.04,0.0,0.0,0.0

jacob,0.87,0.86,0.81,0.76,0.55,0.08,0.0

mse,0.93,0.78,0.31,0.08,0.01,0.0,0.0

mseMargin,0.93,0.82,0.4,0.17,0.01,0.0,0.0

loss2,0.93,0.91,0.89,0.87,0.78,0.51,0.28

loss1,0.94,0.91,0.77,0.59,0.34,0.21,0.12

adv0.1,0.91,0.9,0.87,0.83,0.66,0.01,0.0

adv0.2,0.92,0.77,0.68,0.62,0.49,0.08,0.0

adv0.3,0.94,0.84,0.34,0.16,0.16,0.03,0.0

\end{filecontents*}
\csvautotabular{PRECCNN1100pgd.csv}
\end{center}
\end{table}

%MLP spsa
\begin{table}[H]
\caption{ACC MLP SPSA}
\label{ACC MLP1spsa}
\begin{center}
\begin{filecontents*}{ACCMLP1spsa.csv}
noise level,0,0.01,0.03,0.05,0.1,0.2,0.3

ce,0.93,0.67,0.17,0.02,0.0,0.0,0.0

jacob,0.85,0.82,0.76,0.69,0.42,0.04,0.0

mse,0.91,0.71,0.31,0.19,0.06,0.04,0.03

mseMargin,0.91,0.77,0.45,0.32,0.12,0.02,0.0

loss2,0.9,0.82,0.73,0.67,0.51,0.32,0.2

loss1,0.92,0.87,0.69,0.39,0.08,0.02,0.0

adv0.1,0.91,0.77,0.69,0.69,0.58,0.04,0.0

adv0.2,0.89,0.6,0.52,0.52,0.52,0.23,0.04

adv0.3,0.91,0.59,0.32,0.31,0.32,0.25,0.09

\end{filecontents*}
\csvautotabular{ACCMLP1spsa.csv}
\end{center}
\end{table}

\begin{table}[H]
\caption{PREC MLP SPSA}
\label{PREC MLP1  spsa}
\begin{center}
\begin{filecontents*}{PRECMLP1spsa.csv}
noise level,0,0.01,0.03,0.05,0.1,0.2,0.3

ce,0.93,0.68,0.2,0.05,0.0,0.0,0.0

jacob,0.86,0.83,0.77,0.7,0.46,0.06,0.0

mse,0.93,0.78,0.35,0.22,0.05,0.02,0.01

mseMargin,0.93,0.84,0.54,0.38,0.19,0.02,0.01

loss2,0.92,0.86,0.79,0.73,0.56,0.3,0.19

loss1,0.92,0.88,0.72,0.44,0.08,0.03,0.01

adv0.1,0.92,0.78,0.7,0.71,0.59,0.05,0.0

adv0.2,0.91,0.62,0.54,0.55,0.55,0.27,0.06

adv0.3,0.92,0.62,0.36,0.34,0.36,0.29,0.13

\end{filecontents*}
\csvautotabular{PRECMLP1spsa.csv}
\end{center}
\end{table}

% CNN1 spsa
\begin{table}[H]
\caption{ACC CNN SPSA}
\label{ACC CNN1 spsa}
\begin{center}
\begin{filecontents*}{ACCCNN1spsa.csv}
noise level,0,0.01,0.03,0.05,0.1,0.2,0.3

ce,0.94,0.74,0.14,0.01,0.0,0.0,0.0

jacob,0.85,0.83,0.79,0.75,0.52,0.07,0.0

mse,0.92,0.63,0.18,0.06,0.02,0.01,0.0

mseMargin,0.92,0.75,0.31,0.09,0.01,0.0,0.0

loss2,0.93,0.89,0.81,0.77,0.58,0.22,0.15

loss1,0.93,0.9,0.72,0.52,0.31,0.16,0.05

adv0.1,0.91,0.89,0.86,0.82,0.66,0.01,0.0

adv0.2,0.92,0.73,0.63,0.6,0.6,0.28,0.03

adv0.3,0.94,0.81,0.32,0.15,0.14,0.17,0.15

\end{filecontents*}
\csvautotabular{ACCCNN1spsa.csv}
\end{center}
\end{table}

\begin{table}[H]
\caption{PREC CNN SPSA}
\label{PREC CNN1  spsa}
\begin{center}
\begin{filecontents*}{PRECCNN1spsa.csv}
noise level,0,0.01,0.03,0.05,0.1,0.2,0.3

ce,0.95,0.76,0.19,0.02,0.0,0.0,0.0

jacob,0.86,0.84,0.8,0.75,0.53,0.1,0.0

mse,0.93,0.72,0.24,0.06,0.02,0.0,0.0

mseMargin,0.93,0.8,0.37,0.11,0.0,0.0,0.0

loss2,0.94,0.91,0.86,0.83,0.74,0.2,0.07

loss1,0.93,0.91,0.75,0.54,0.29,0.16,0.05

adv0.1,0.91,0.9,0.87,0.83,0.68,0.03,0.0

adv0.2,0.92,0.76,0.66,0.63,0.63,0.32,0.02

adv0.3,0.94,0.83,0.32,0.16,0.16,0.21,0.17

\end{filecontents*}
\csvautotabular{PRECCNN1spsa.csv}
\end{center}
\end{table}

\section{Appendix}

In this section, ECG signals with different levels of noise are shown. The attack is PGD-100 and the target network is MLP, which has been trained with Cross-entropy loss for 50 epochs. As we can see, when the noise level reaches 0.1 (shown in Figure 8 (f)), the ECG signal is even hardly recognizable by human eye.

\begin{figure}[h]
\begin{center}

\begin{subfigure}{0.45\textwidth}
\includegraphics[width=0.9\linewidth, height=5cm]{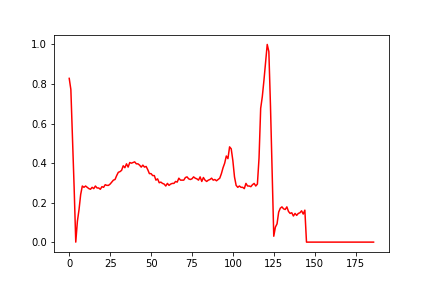} 
\caption{clean ECG}
\label{1}
\end{subfigure}
\begin{subfigure}{0.45\textwidth}
\includegraphics[width=0.9\linewidth, height=5cm]{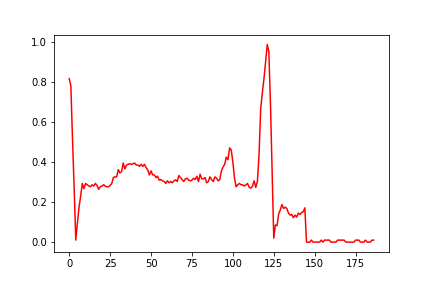}
\caption{ECG on noise level 0.01}
\label{2}
\end{subfigure}

\begin{subfigure}{0.45\textwidth}
\includegraphics[width=0.9\linewidth, height=5cm]{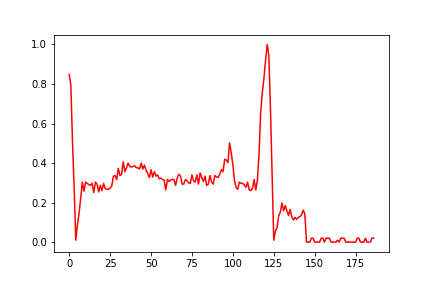}
\caption{ECG on noise level 0.02}
\label{3}
\end{subfigure}
\begin{subfigure}{0.45\textwidth}
\includegraphics[width=0.9\linewidth, height=5cm]{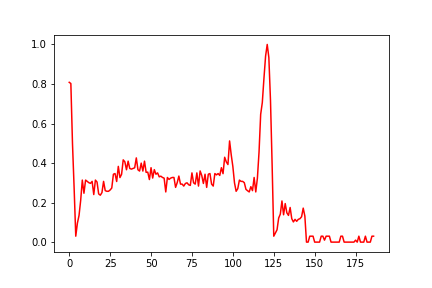}
\caption{ECG on noise level 0.03}
\label{4}
\end{subfigure}

\begin{subfigure}{0.45\textwidth}
\includegraphics[width=0.9\linewidth, height=5cm]{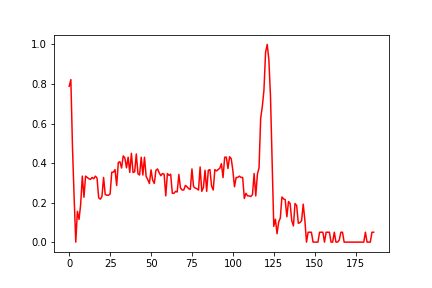}
\caption{ECG on noise level 0.05}
\label{5}
\end{subfigure}
\begin{subfigure}{0.45\textwidth}
\includegraphics[width=0.9\linewidth, height=5cm]{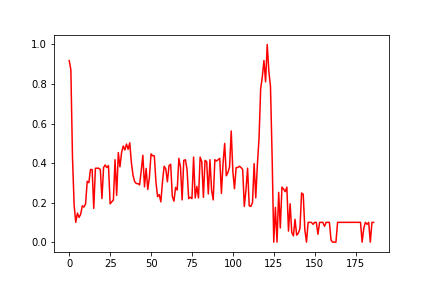}
\caption{ECG on noise level 0.1}
\label{6}
\end{subfigure}

\begin{subfigure}{0.45\textwidth}
\includegraphics[width=0.9\linewidth, height=5cm]{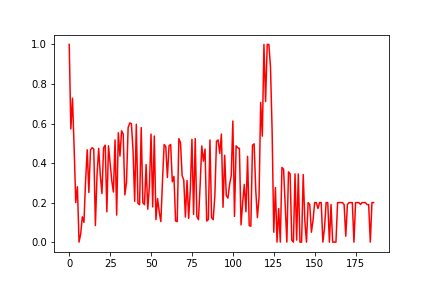}
\caption{ECG on noise level 0.2}
\label{7}
\end{subfigure}
\begin{subfigure}{0.45\textwidth}
\includegraphics[width=0.9\linewidth, height=5cm]{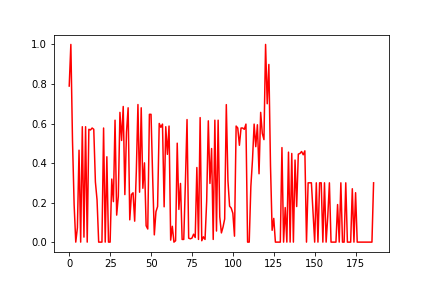}
\caption{ECG on noise level 0.3}
\label{8}
\end{subfigure}
\end{center}
\caption{ECG signals on different noise levels}
\end{figure}

\section{Appendix}
In the experiment, loss funtions $loss_1$ and $loss_2$ run in mini-batches. Assume there are $k$ classes,$loss_1$ should be:

\begin{equation}
R_1= \sum_i || w_i - \gamma \frac {X}{X^T X}||_2^2
\end{equation}

We combine the regularization term $R_1$ with mean square error (MSE) loss and margin loss for classification, given by

\begin{equation}
loss_1(X,Y)=\sum (Z-Y)^2+multi\_margin\_loss(Y, Z) + \beta_1 R_1
\end{equation}

$loss_2$ should be:

\begin{equation}
R_2 = \sum_i^k \frac{||w_i||_1 . \epsilon_{max}}{|z_i|} 
\end{equation}

We combine the regularization term $R_2$ with mean square error (MSE) loss and margin loss for classification, given by

\begin{equation}
loss_2 (X, Y)=(Z-Y)^2+multi\_margin\_loss(Y, Z)+\beta_2 log(1+R_2)
\end{equation}

\end{document}

%% file: math_commands.tex
\usepackage{amsmath,amsfonts,bm}

\def\eqref#1{equation~\ref{#1}}
\def\1{\bm{1}}

\DeclareMathAlphabet{\mathsfit}{\encodingdefault}{\sfdefault}{m}{sl}
\SetMathAlphabet{\mathsfit}{bold}{\encodingdefault}{\sfdefault}{bx}{n}

%% file: iclr2020_conference.bbl
\begin{thebibliography}{13}
\providecommand{\natexlab}[1]{#1}
\providecommand{\url}[1]{\texttt{#1}}
\expandafter\ifx\csname urlstyle\endcsname\relax
  \providecommand{\doi}[1]{doi: #1}\else
  \providecommand{\doi}{doi: \begingroup \urlstyle{rm}\Url}\fi

\bibitem[ECG()]{ECG}
{ECG Heartbeat Categorization Dataset}.
\newblock URL \url{https://www.kaggle.com/shayanfazeli/heartbeat}.

\bibitem[Akhtar \& Mian(2018)Akhtar and Mian]{Akhtar2018}
Naveed Akhtar and Ajmal Mian.
\newblock {Threat of Adversarial Attacks on Deep Learning in Computer Vision: A
  Survey}.
\newblock \emph{IEEE Access}, 6\penalty0 (August):\penalty0 14410--14430, 2018.
\newblock ISSN 21693536.
\newblock \doi{10.1109/ACCESS.2018.2807385}.

\bibitem[Ding et~al.(2018)Ding, Sharma, Lui, and Huang]{Ding2018}
Gavin~Weiguang Ding, Yash Sharma, Kry Yik~Chau Lui, and Ruitong Huang.
\newblock {Max-Margin Adversarial (MMA) Training: Direct Input Space Margin
  Maximization through Adversarial Training}.
\newblock 01\penalty0 (1):\penalty0 1--34, 2018.
\newblock URL \url{http://arxiv.org/abs/1812.02637}.

\bibitem[Goodfellow et~al.(2015)Goodfellow, Shlens, and
  Szegedy]{Goodfellow2015}
Ian~J. Goodfellow, Jonathon Shlens, and Christian Szegedy.
\newblock {Explaining and harnessing adversarial examples}.
\newblock \emph{3rd International Conference on Learning Representations, ICLR
  2015 - Conference Track Proceedings}, pp.\  1--11, 2015.

\bibitem[Hannun et~al.(2019)Hannun, Rajpurkar, Haghpanahi, Tison, Bourn,
  Turakhia, and Ng]{hannun2019cardiologist}
Awni~Y Hannun, Pranav Rajpurkar, Masoumeh Haghpanahi, Geoffrey~H Tison, Codie
  Bourn, Mintu~P Turakhia, and Andrew~Y Ng.
\newblock Cardiologist-level arrhythmia detection and classification in
  ambulatory electrocardiograms using a deep neural network.
\newblock \emph{Nature medicine}, 25\penalty0 (1):\penalty0 65, 2019.

\bibitem[Jakubovitz \& Giryes(2018)Jakubovitz and
  Giryes]{jakubovitz2018improving}
Daniel Jakubovitz and Raja Giryes.
\newblock Improving dnn robustness to adversarial attacks using jacobian
  regularization.
\newblock In \emph{Proceedings of the European Conference on Computer Vision
  (ECCV)}, pp.\  514--529, 2018.

\bibitem[Kachuee et~al.(2018)Kachuee, Fazeli, and Sarrafzadeh]{Kachuee2018}
Mohammad Kachuee, Shayan Fazeli, and Majid Sarrafzadeh.
\newblock {ECG heartbeat classification: A deep transferable representation}.
\newblock \emph{Proceedings - 2018 IEEE International Conference on Healthcare
  Informatics, ICHI 2018}, pp.\  443--444, 2018.
\newblock \doi{10.1109/ICHI.2018.00092}.

\bibitem[Madry et~al.(2017)Madry, Makelov, Schmidt, Tsipras, and
  Vladu]{madry2017towards}
Aleksander Madry, Aleksandar Makelov, Ludwig Schmidt, Dimitris Tsipras, and
  Adrian Vladu.
\newblock Towards deep learning models resistant to adversarial attacks.
\newblock \emph{arXiv preprint arXiv:1706.06083}, 2017.

\bibitem[Moody \& Mark(2001)Moody and Mark]{moody2001impact}
George~B Moody and Roger~G Mark.
\newblock The impact of the mit-bih arrhythmia database.
\newblock \emph{IEEE Engineering in Medicine and Biology Magazine}, 20\penalty0
  (3):\penalty0 45--50, 2001.

\bibitem[Papernot et~al.(2017)Papernot, McDaniel, Goodfellow, Jha, Celik, and
  Swami]{papernot2017practical}
Nicolas Papernot, Patrick McDaniel, Ian Goodfellow, Somesh Jha, Z~Berkay Celik,
  and Ananthram Swami.
\newblock Practical black-box attacks against machine learning.
\newblock In \emph{Proceedings of the 2017 ACM on Asia conference on computer
  and communications security}, pp.\  506--519, 2017.

\bibitem[Ros \& Doshi-velez(2017)Ros and Doshi-velez]{Ros2017}
Andrew~Slavin Ros and Finale Doshi-velez.
\newblock {Improving the Adversarial Robustness and Interpretability of Deep
  Neural Networks by Regularizing Their Input Gradients}.
\newblock pp.\  1660--1669, 2017.

\bibitem[Shafahi et~al.(2019)Shafahi, Najibi, Ghiasi, Xu, Dickerson, Studer,
  Davis, Taylor, and Goldstein]{shafahi2019adversarial}
Ali Shafahi, Mahyar Najibi, Mohammad~Amin Ghiasi, Zheng Xu, John Dickerson,
  Christoph Studer, Larry~S Davis, Gavin Taylor, and Tom Goldstein.
\newblock Adversarial training for free!
\newblock In \emph{Advances in Neural Information Processing Systems}, pp.\
  3353--3364, 2019.

\bibitem[Uesato et~al.(2018)Uesato, O'Donoghue, Oord, and
  Kohli]{uesato2018adversarial}
Jonathan Uesato, Brendan O'Donoghue, Aaron van~den Oord, and Pushmeet Kohli.
\newblock Adversarial risk and the dangers of evaluating against weak attacks.
\newblock \emph{arXiv preprint arXiv:1802.05666}, 2018.

\end{thebibliography}
